\documentclass[11pt,3p,review,authoryear,dvipsnames]{elsarticle}

\usepackage[table,xcdraw]{xcolor}

\usepackage{float}
\usepackage{enumitem}
\usepackage{hyperref}
\hypersetup{
colorlinks=true,
linkcolor=blue,
filecolor=blue,
urlcolor=blue,
pdftitle={Overleaf Example},
pdfpagemode=FullScreen,
citecolor=blue,
}

\usepackage{listings}
\usepackage{xurl}
\usepackage{amsmath,bm}
\usepackage{graphicx}
\usepackage[bottom]{footmisc}
\hypersetup{
    colorlinks=true,
    linkcolor=blue,
    filecolor=magenta,      
    urlcolor=blue,
}
\usepackage{booktabs}
\usepackage{multirow}
\usepackage[nameinlink]{cleveref}
\usepackage{subcaption}
\usepackage{amssymb}
\usepackage{pdfpages}

\journal{arXiv}
\biboptions{longnamesfirst}

\begin{document}

\begin{frontmatter}

\title{Predicting the Price Movement of Cryptocurrencies Using Linear Law-based Transformation}

\author[1,2]{Marcell T. Kurbucz\corref{cor1}}
\cortext[cor1]{Corresponding author}
\ead{kurbucz.marcell@wigner.hu}
\author[1]{P\'eter P\'osfay}
\ead{posfay.peter@wigner.hu}
\author[1]{Antal Jakov\'ac}
\ead{jakovac.antal@wigner.hu}

\address[1]{Department of Computational Sciences, Wigner Research Centre for Physics, 29-33 Konkoly-Thege Mikl\'os Street, \\ H-1121 Budapest, Hungary}
\address[2]{Institute of Data Analytics and Information Systems, Corvinus University of Budapest, 8 F\H{o}v\'am Square, \\ H-1093 Budapest, Hungary}

\begin{abstract}
The aim of this paper is to investigate the effect of a novel method called linear law-based feature space transformation (LLT) on the accuracy of intraday price movement prediction of cryptocurrencies. To do this, the 1-minute interval price data of Bitcoin, Ethereum, Binance Coin, and Ripple between 1 January 2019 and 22 October 2022 were collected from the Binance cryptocurrency exchange. Then, $14$-hour nonoverlapping time windows were applied to sample the price data. The classification was based on the first $12$ hours, and the two classes were determined based on whether the closing price rose or fell after the next $2$ hours. These price data were first transformed with the LLT, then they were classified by traditional machine learning algorithms with $10$-fold cross-validation. Based on the results, LLT greatly increased the accuracy for all cryptocurrencies, which emphasizes the potential of the LLT algorithm in predicting price movements.
\end{abstract}


\end{frontmatter}

\section{Introduction}
\label{S:1}

The advent of cryptocurrencies has revolutionized the world of finance and investment.\footnote{The concept of Bitcoin, i.e. the first blockchain application, was made public in \citet{nakamoto2008bitcoin}.} Cryptocurrencies, such as Bitcoin and Ethereum, are decentralized digital assets that operate on blockchain technology. Contrary to traditional financial systems, this technology ignores financial intermediaries and provides a public ledger that records all transactions. Predicting the price movement of these currencies creates a number of challenges, especially due to the extremely high volatility of their exchange prices \citep{mahayana2022predicting}. As reflected by current market dynamics, the absence of government regulation and oversight creates obstacles in mitigating the high volatility and consequential losses that may arise \citep{dag2023tree}. For this reason, traders and investors must use appropriate price prediction methods that consider the extreme behavior of cryptocurrency markets \citep{mba2020markov}.

In recent years, several studies have been conducted on the classification of intraday price movements of cryptocurrencies. For instance, \citet{el2021adaptive} proposed a deep learning model for forecasting and classifying the price of various cryptocurrencies based on a multiple-input architecture. To identify the relevant variables, they applied a two-stage adaptive feature selection procedure. \citet{mahayana2022predicting} predicted the price movement of Bitcoin's exchange rate using a tree-based classification algorithm with the gradient boosting framework. In \citet{zhou2023multi}, price movements were predicted by a support vector machine algorithm based on historical trading data, sentiment indicators, and daily Google Trends. Additionally, a data-driven tree augmented na\"{i}ve Bayes methodology was proposed by \citet{dag2023tree} that can be used for identifying the most important factors influencing the price movements of Bitcoin. Other works focus on the predictive power of the features obtained from the transaction network of various cryptocurrencies \citep{abay2019chainnet,akcora2018forecasting,akcora2018bitcoin,dey2020role,kurbucz2019predicting}.

The recently published algorithm called linear law-based feature space transformation (LLT) \citep{kurbucz2022facilitating} can be applied to facilitate uni- and multivariate time series classification tasks. The aim of this paper is to investigate the effect of LLT on the accuracy of intraday price movement prediction of various cryptocurrencies. To do this, the 1-minute interval price data of Bitcoin, Ethereum, Binance Coin, and Ripple between 1 January 2019 and 22 October 2022 were collected from the Binance cryptocurrency exchange. Then, $14$-hour nonoverlapping time windows were applied to sample the price data. The classification was based on the first $12$ hours, and the two classes were determined based on whether the closing price rose or fell after the next $2$ hours. These price data were first transformed with the LLT, and then they were classified by traditional machine learning algorithms with $10$-fold cross-validation and Bayesian hyperparameter optimization.

The rest of this paper is organized as follows. \Cref{S:2} introduces the employed dataset, the classification task, the LLT algorithm, and the applied software and its settings. \Cref{S:3} compares and discusses the classification outcomes obtained with and without the LLT. Finally, conclusions and suggested future research directions are provided in \Cref{S:3}.

\section{Data and methodology}
\label{S:2}

\subsection{Cryptocurrency dataset}
\label{SS:2.1}
\noindent

This study is based on the 1-minute interval price data of Bitcoin (BTC), Ethereum (ETH), Binance Coin (BNB), and Ripple (XRP) between 1 January 2019 and 22 October 2022. These data were collected from the Binance cryptocurrency exchange by the CryptoDataDownload website (\url{https://www.cryptodatadownload.com/}, retrieved: 19 April 2023). For each cryptocurrency, the obtained dataset contains the opening, closing, high, and low prices, as well as the transaction volume and the volume of Tether (USDT): i.e., the volume of the stablecoin that prices were measured against. Hereinafter, these variables are called the initial features of cryptocurrencies.

\subsection{Classification task}
\label{SS:2.2}

\begin{figure}[hbt!]
 \centering
  \includegraphics[width=0.55\textwidth]{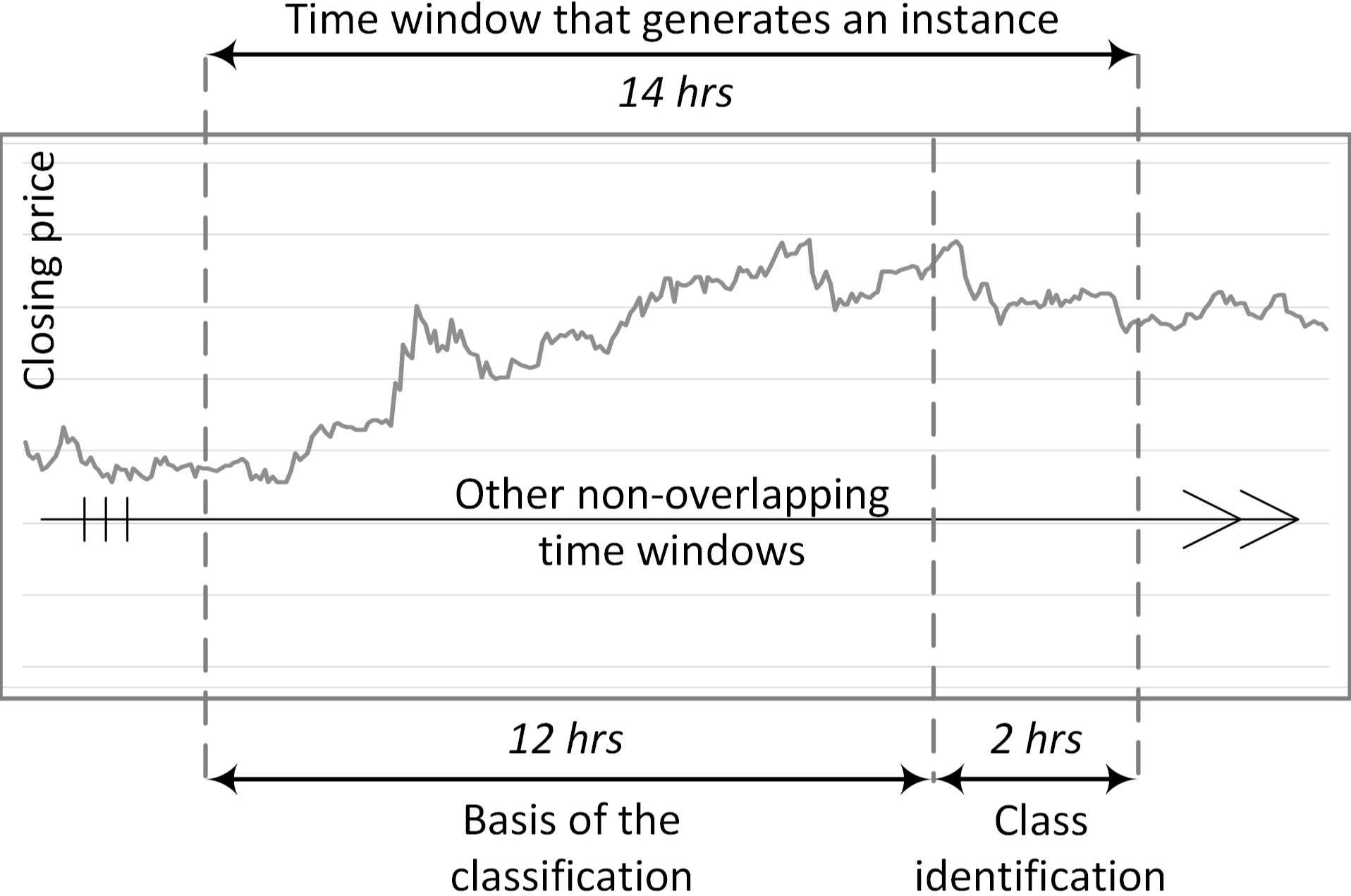}
\caption{Applied sampling procedure}
\label{F:1}
\end{figure}

To define the classification task, we first generated instances by sampling the price datasets based on $14$-hour, nonoverlapping time windows. The input data ($\bm X$) of the classification task were the $720$ ($k$) consecutive values of the $6$ ($m$) initial features, measured in the first $12$ hours. The output variable contains two classes defined by whether the closing price rose or fell after the next $2$ hours. After we balanced the number of instances related to the two classes, this sampling procedure resulted in approximately $1\,680$ ($n$) instances in each cryptocurrency: $840$ per class. The applied sampling procedure is illustrated in Fig. \ref{F:1}.

Formally, input data are denoted by $\bm X = \{ \bm X_t \;\vert \; t\in \{1,2,\dots,k\}\}$, where $t$ represents the observation times. The internal structure of the input data is $\bm X_t = \{ \bm x_t^{i,j} \;\vert\; i\in\{1,2,\dots,n\},~j\in \{1,2,\dots,m\}\}$, where $i$ indicates the instances and $j$ identifies the different initial features belonging to a given instance. The output is a vector $\bm y \in \{0,1\}$ identifying the classes of instances ($ \bm y = \{ y^{i}\in\mathbb{R}\;\vert\; i\in\{1,2,\dots,n\}\}$). The goal of the classification task is to predict the $\bm{y}$ values (classes) from the $\bm{X}$ input data.

\subsection{Price movement prediction}
\label{SS:2.3}

Price movement prediction is based on the combination of the LLT with traditional machine learning algorithms.\footnote{The concept of linear laws is detailed in \citet{Jakovac_time_series} and \citet{jakovac2022reconstruction}. The complete mathematical formulation of LLT can be found in \citet{kurbucz2022facilitating}.}

LLT first separates the ($tr\in\small\{1,2,\dots,\tau\small\}$) and tests ($te\in\small\{\tau+1,\tau+2,\dots,n\small\}$) the sets of instances.\footnote{Instances are split in such a way that their classes are balanced in the two sets. For transparency, we assume that the original arrangement of the instances in the dataset satisfies this condition for the $tr$ and $te$ sets.} Then, it identifies the governing patterns (linear laws) of each input sequence in the training set. To this end, we perform the $l^{\text{th}}$ order ($l \in \mathbb{Z}^+$ and $l<k$) time-delay embedding \citep{takens1981dynamical} of these series as follows:

\begin{equation}
\label{eq:A_matrix}
\bm{A}^{tr,j}=\left(\begin{matrix}\bm{x}^{tr,j}_{1}&\bm{x}^{tr,j}_{2}&\cdots&\bm{x}^{tr,j}_{l}\\\
\bm{x}^{tr,j}_{2}&\ddots&\ddots&\vdots\\\vdots&\ddots&\ddots&\vdots\\\bm{x}^{tr,j}_{k-l}&\cdots&\cdots&\bm{x}^{tr,j}_{k}\\\end{matrix}\right),
\end{equation}

\noindent
for all $tr$ and $j$. Then, symmetric $l \times l$ matrices are generated as $\bm{S}^{tr,j}={\bm{A}^{tr,j}}^\intercal\bm{A}^{tr,j}$. The coefficients ($\bm{v}^{tr,j}$) that satisfy $\bm{S}^{tr,j} \bm{v}^{tr,j} \approx \mathbf{0}$ are called the linear law of the input series $\bm{x}^{tr,j}_t$. These laws are grouped by input series and classes as follows: $\bm{V}^j = \{\bm{V}^j_{0},\bm{V}^j_{1}\}$, where $\bm{V}^j_{0}$ and $\bm{V}^j_{1}$ denote the laws of the training set related to the initial feature $j$ and the two classes.

The next step involves calculating $\bm{S}^{te,j}$ matrices from the test instance's initial features and left multiplying them by the $\bm{V}^j$ matrices obtained from the same initial features of the training instances ($\bm{S}^{\tau+1,1}\bm{V}^1,\bm{S}^{\tau+1,2}\bm{V}^2,\dots,\bm{S}^{n,m}\bm{V}^m$). The product of these matrices gives an estimate of whether the test data belong to the same class as the training instance based on the presence of close-to-null vectors with a small variance in the resulting matrix. The final step reduces the dimensions of the resulting matrices by selecting columns with the smallest variance and/or absolute mean from the $\bm{S}^{te,j}\bm{V}^j$ matrices for each class. As a result of this step, the transformed feature space of the test set has $((n-\tau) l) \times (m c + 1)$ dimensions combined with the output variable.

After the feature space transformation, test instances are classified by decision tree (DT) \citep{li2019discovering}, k-nearest neighbor (KNN) \citep{mello2019time}, support vector machine (SVM) \citep{altobi2019fault,raghu2019performance}, and classifier ensemble (ENS) \citep{oza2008classifier} algorithms with cross-validation and Bayesian hyperparameter optimization.

\subsection{Applied software and settings}
\label{SS:2.4}

Datasets were transformed using the \texttt{LLT} R package (version: 0.1.0) \citep{kurbucz2023plos} with the following settings:\footnote{The \texttt{LLT} R package is publicly available on GitHub \citep{git_llt}.}

\begin{itemize}[noitemsep]
\item \texttt{test\_ratio = 0.25}: $25\%$ of the instances were included in the test set. The training and test sets contained approximately $1284$ ($\tau$) and $428$ ($n-\tau$) independent instances for each cryptocurrency: $642$ and $214$ from both classes, respectively.
\item \texttt{dim = 10}: It defined the $l$ parameter ($l = 10$).
\item \texttt{lag = 11}: The successive row lag of the $\bm{A}$ matrix is set to $11$. That is, since $l = 10$, $\bm{A}^{tr,j}_{1,10} = \bm{x}^{tr,j}_{10}$, $\bm{A}^{tr,j}_{2,1} = \bm{x}^{tr,j}_{11}$, $\bm{A}^{tr,j}_{2,2} = \bm{x}^{tr,j}_{12}$, and so on.
\item \texttt{select = "var"}: As the last step of the LLT, the column vectors with the smallest variance were selected from the $\bm{S}^{te,j}\bm{V}^j$ matrices for each class.
\item \texttt{seed = 12345}: For the reproducibility of the results, the random number generation was fixed.
\end{itemize}

The original and transformed classification tasks were solved in the Classification Learner App of MATLAB.\footnote{More information can be found at \url{https://www.mathworks.com/help/stats/classificationlearner-app.html}, retrieved: 19 April 2023).} For each classifier, $10$-fold cross-validation and $300$-step Bayesian hyperparameter optimization were applied.

\section{Results and discussion}
\label{S:3}

The results of the original and transformed classification tasks are presented in Table \ref{tab:results}.\footnote{Related confusion matrices and the details of hyperparameter optimization can be found in the \nameref{s:sm}.}

\begin{table}[H]
\caption{Classification accuracies (\%)}
\centering
\label{tab:results}
    \begin{subtable}{.40\textwidth}
    \centering
    \caption{Original feature space}
    \resizebox{\textwidth}{!}{
    \begin{tabular}{@{}lcccc}
                  & \multicolumn{1}{c}{\textbf{BTC}} & \multicolumn{1}{c}{\textbf{ETH}} & \multicolumn{1}{c}{\textbf{BNB}} & \multicolumn{1}{c}{\textbf{XRP}} \\\hline \hline
        \textbf{Ensemble} &     56.5                     &     55.8                     &   52.3                                &  56.2                                 \\
        \textbf{KNN}      &            57.0                   &         55.8                 &   57.5                                 &   57.3                                \\
        \textbf{DT}       &             57.0                        &          54.2                      &     50.8                                 &  52.4                                 \\
        \textbf{SVM}      &     59.6                        &           56.1                   &   57.5                                 & 54.5                                  \\ \hline
        \end{tabular}
        }
    \end{subtable}
 \hspace{1em}
    \begin{subtable}{.395\textwidth}
      \centering
      \caption{\centering Feature space transformed by LLT}
    \resizebox{\textwidth}{!}{%
    \begin{tabular}{@{}lcccc}
                  & \multicolumn{1}{c}{\textbf{BTC}} & \multicolumn{1}{c}{\textbf{ETH}} & \multicolumn{1}{c}{\textbf{BNB}} & \multicolumn{1}{c}{\textbf{XRP}} \\\hline \hline
        \textbf{Ensemble} &     75.2                     &    80.8                     &   70.4                                &  79.5                                 \\
        \textbf{KNN}      &            84.3                   &        82.0                 &   77.6                                &   81.4                                \\
        \textbf{DT}       &             65.9                        &          73.6                      &     60.8                                  &  67.5                                 \\
        \textbf{SVM}      &     65.9                        &          64.3                   &   58.8                                 & 62.0                                  \\ \hline
        \end{tabular}
    }
    \end{subtable}
\end{table}

As shown in Table \ref{tab:results}, LLT greatly increased the accuracy for all classifiers and cryptocurrencies. In the case of the original feature space, the SVM algorithm achieved the best average performance with an accuracy of $56.9$\%. After the transformation, the KNN algorithm became the most accurate classifier, with an average accuracy of $81.3$\%. This result is consistent with our previous work \citep{kurbucz2022facilitating}, in which we tested the same classifiers on human activity recognition data and found that the combination of LLT and KNN achieved the highest accuracy and shortest computation time, outperforming even state-of-the-art methods. Since LLT has a low calculation requirement, it can effectively handle feature spaces with much higher dimensions than previously used methods. Applying additional features, such as sentiment indicators and daily Google Trends \citep{zhou2023multi}, can result in even higher classification accuracy.

\section{Conclusions and future works}
\label{S:4}

This paper investigated the effect of LLT on the accuracy of intraday price movement prediction of cryptocurrencies. To do this, the 1-minute interval price data of Bitcoin, Ethereum, Binance Coin, and Ripple between 1 January 2019 and 22 October 2022 were collected from the Binance cryptocurrency exchange. Then, $14$-hour nonoverlapping time windows were applied to sample the price data. The classification was based on the first $12$ hours, and the two classes were determined based on whether the closing price rose or fell after the next $2$ hours. These price data were first transformed with the LLT and then classified by traditional machine learning algorithms with $10$-fold cross-validation.

Based on the results, LLT greatly increased the accuracy regardless of the type of cryptocurrency and classification algorithm. While the SVM algorithm achieved the best results for the original feature space, after the transformation, we achieved the highest average accuracy with the KNN classifier. By using the LLT algorithm, we managed to increase the best average accuracy from $56.9$\% to $81.3$\%. These results not only emphasize the potential of the LLT algorithm in price movement prediction but also provide further research directions. Future works could focus on the classification performance of the LLT-KNN algorithm pair in high-dimensional feature spaces. Other research could extend the LLT algorithm with the adaptive selection of the laws used during the transformation. Finally, due to its low computational cost, LLT could also be useful in the field of portfolio optimization, which would require further investigation.

\section*{Data availability}

The applied price data were collected from the Binance cryptocurrency exchange by the CryptoDataDownload website (\url{https://www.cryptodatadownload.com/}, retrieved: 19 April 2023).

\section*{Supplementary material}
\label{s:sm}

The supplementary material contains the confusion matrices related to Table \ref{tab:results} and the results of hyperparameter optimization applied during the calculations.

\section*{Acknowledgements}
Project no. PD142593 was implemented with the support provided by the Ministry of Culture and Innovation of Hungary from the National Research, Development, and Innovation Fund, financed under the PD\_22 ``OTKA'' funding scheme. A.J. received support from the Hungarian Scientific Research Fund (OTKA/NRDI Office) under contract number K123815. The research was supported by the Ministry of Innovation and Technology NRDI Office within the framework of the MILAB Artificial Intelligence National Laboratory Program.

\bibliography{refs}

\newpage
\includepdf[pages=-]{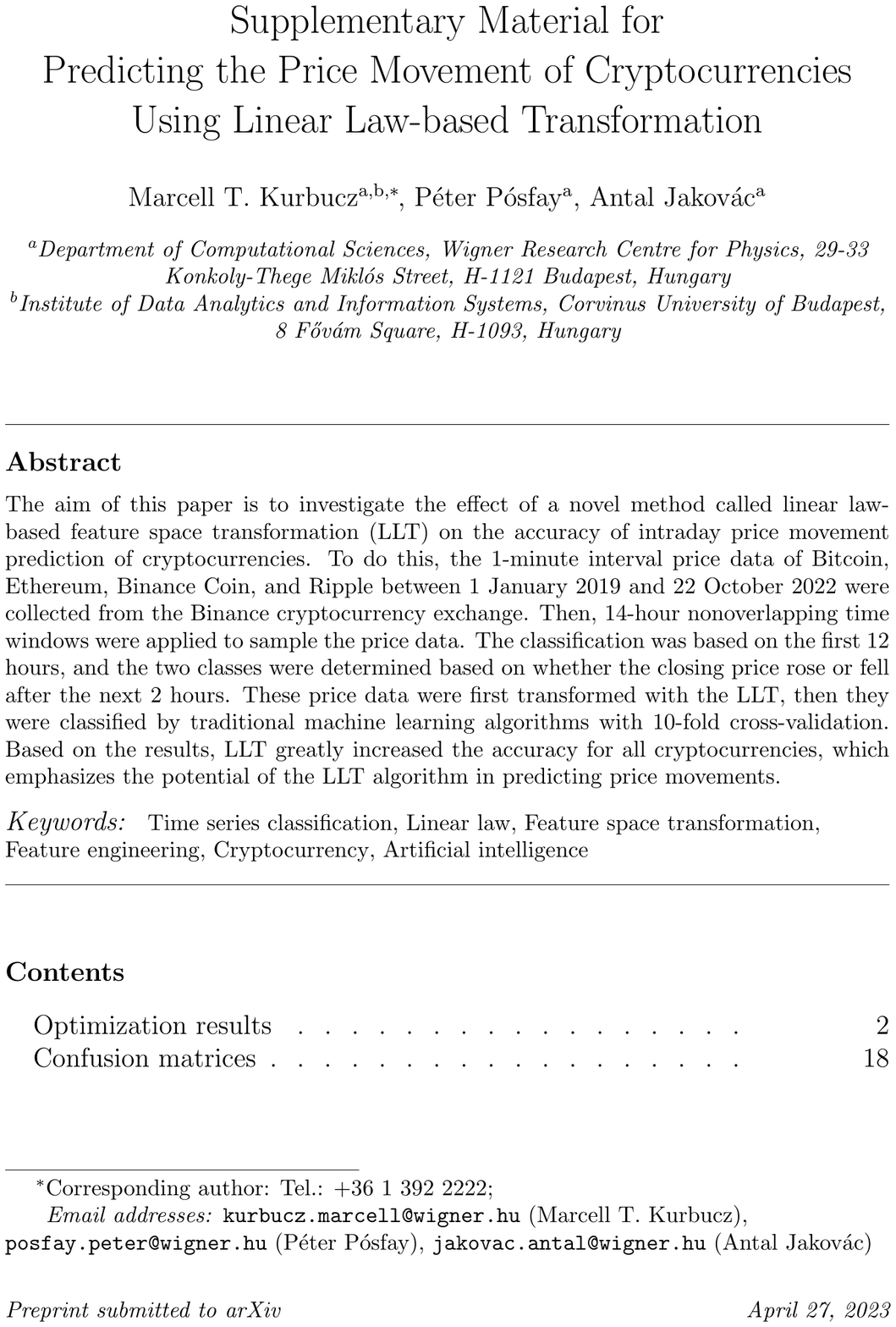}

\end{document}